\def\year{2020}\relax
\documentclass[letterpaper]{article} % DO NOT CHANGE THIS
\usepackage{aaai20}  % DO NOT CHANGE THIS
\usepackage{times}  % DO NOT CHANGE THIS
\usepackage{helvet} % DO NOT CHANGE THIS
\usepackage{courier}  % DO NOT CHANGE THIS
\usepackage[hyphens]{url}  % DO NOT CHANGE THIS
\usepackage{graphicx} % DO NOT CHANGE THIS
\usepackage{hyperref}
\usepackage{subfigure}
\usepackage{color}
\usepackage{balance}
\usepackage{footnote}
\usepackage{tablefootnote}
\newcommand{\red}[1]{\textcolor{red}{#1}}
\newcommand{\blue}[1]{\textcolor{blue}{#1}}

\newcounter{JRNumberOfComments}
\stepcounter{JRNumberOfComments}

\newcounter{SSGNumberOfComments}
\stepcounter{SSGNumberOfComments}

\title{A Dataset of Fact-Checked Images \\Shared on WhatsApp During the Brazilian and Indian Elections*}

\author{
\Large\textbf{{Julio C. S. Reis$^{\star}$, Philipe Melo$^{\star}$, Kiran Garimella$^{\ddag}$,}} \\
\Large\textbf{{Jussara M. Almeida$^{\star}$, Dean Eckles$^{\ddag}$, Fabr\'icio Benevenuto$^{\star}$}}\\
	$^{\star}$Computer Science Department, Universidade Federal de Minas Gerais (UFMG), Brazil\\
   $^{\ddag}$Massachusetts Institute of Technology (MIT), United States\\
    {\{julio.reis, philipe, jussara, fabricio\}@dcc.ufmg.br}, 
	{\{garimell, eckles\}@mit.edu}\\ 
}

\begin{document}
\maketitle
\begin{abstract}
Recently, messaging applications, such as WhatsApp, have been reportedly abused by misinformation campaigns, especially in Brazil and India. A notable form of abuse in WhatsApp relies on several manipulated images and memes containing all kinds of fake stories. 
In this work, we performed an extensive data collection from a large set of WhatsApp publicly accessible groups and  fact-checking agency websites. This paper opens a novel dataset to the research community containing fact-checked fake images shared through WhatsApp for two distinct scenarios known for the  spread of fake news on the platform: the 2018 Brazilian elections and the 2019 Indian elections. 
\end{abstract}

\section{Introduction}

Digital platforms, including social media systems and messaging applications, have become key environments for news dissemination. Despite the numerous benefits that these systems bring to our society, they have been reportedly abused by misinformation especially for political purposes~\cite{bessi2016social,ribeiro2019@fat,resendeWWW19}.

Election after election, we have seen different forms of abuse and complex strategies of opinion manipulation through the spread of misinformation. The 2016 presidential elections in the USA are still remembered for a `misinformation war' that happened mostly through Twitter and Facebook. The notorious case involved an attempt of influence from Russia through targeting advertising~\cite{ribeiro2019@fat}. Similar attempts were observed during the 2018 Brazilian elections, where WhatsApp was abused to send out misinformation campaigns, with large use of manipulated images and memes containing all kinds of political attacks. A recent study showed that 88\% of the most popular images shared in the last month before the Brazilian elections were fake or misleading~\cite{nyt2018benevenuto}. 
In India, fake rumors spread through WhatsApp were responsible for multiple cases of lynching and social unrest~\cite{arun2019whatsapp}.   

Broadly speaking, there are two kinds of efforts related to the problem of misinformation dissemination. One aims at understanding the phenomenon and describing common patterns and signatures of a fake news~\cite{vosoughi2018spread,resendeWWW19}. The other consists of attempts to provide a solution to mitigate the problem~\cite{melo2019whatappMonitor}.  In both cases, a key challenge for researchers in the field is the lack of public datasets containing fact-checked content. 
The cost associated to build this kind of dataset if huge, as it requires an investigation by experts and journalists, who debunk the fake story usually by checking out all facts and proofs regarding the topic. 
As an example, Comprova, a large collaborative fact-checking initiative from the First Draft News has brought together journalists from 24 different Brazilian media companies and generated 146 reports from June to October 2018~\cite{firstdraf2019report}.

This work opens a novel dataset to the research community, consisting of two sets of 135 and 897 images containing misinformation from Brazil and India, respectively. 
These images circulated on hundreds of publicly accessible WhatsApp groups\footnote{Whatsapp groups are made effectively publicly accessible when group administrators openly share invitation links on the Web and online social networks.}  around the 2018 Brazilian national elections and the 2019 Indian national elections, and were fact-checked by well-known fact-checking agencies. The dataset can be found in the following link:
\url{http://doi.org/10.5281/zenodo.3779157}.
%\url{http://doi.org/10.5281/zenodo.3734805}.
%\url{https://doi.org/10.5281/zenodo.3609246}.
This data is based on a collection of messages shared on the monitored WhatsApp groups, along with data gathered from fact-checking websites and manual expert annotation. We extract all images from both WhatsApp and fact-checking websites, and matched them to build a dataset that covers fact-checked images widely shared on WhatsApp. This includes two distinct political scenarios from Brazil, and India, where WhatsApp played an important role during their respective elections\footnote{\url{https://www.nytimes.com/2018/05/14/technology/whatsapp-india-elections.html}.}

The rest of the paper is organized as follows: We start by surveying existing open datasets in the field. Then, we describe how we built our dataset, presenting its properties and limitations. Finally, we conclude discussing potential applications and implications of our dataset.

\section{Related Datasets}

To make concrete contributions against misinformation, researchers need a wide and broad set of datasets containing labeled data, i.e., fact-checked content, covering different topics and contexts~\cite{hui2018hoaxy,norregaard2019nela}. In the particular context of misinformation during election campaigns, data covering multiple elections is also of interest to reveal potentially different properties or reinforce common characteristics. 

We executed a brief survey on existing public datasets commonly used by  prior studies on the phenomenon of fake news, aiming at  either understanding it or proposing solutions to mitigate its effects.
Those datasets usually contain content labeled  as fake or true stories. Such fact-checked contents often appears in a range of different formats, such as news articles, claims or quotes by celebrities, rumors, reports, or images. They also cover different scenarios such as wars and politics.  

Table~\ref{tab:datasets} summarizes some of the well-known fact-checked datasets and their main characteristics, including a description, the total number of instances as well as their distribution by label (i.e. fact-checking verdict -- true; fake; etc.) and information about raters (i.e. fact-checkers). Note that we colored in \red{red} the number of fact-checked instances labeled as fake, in \blue{blue}, the true news, and in black the remaining ones (i.e. those that are neutral).

\begin{table*}[!htpb]
  \centering
	 %\scriptsize
	 \footnotesize
	 %\small
    \begin{tabular}{|p{4.1cm}|p{4cm}|p{2.2cm}|p{2cm}|c|}
		\hline
    \textbf{Dataset} & \textbf{Description} & \textbf{Labels} & \textbf{Raters} & \textbf{\# Instances} \\ \hline
    \textbf{BuzzFace}\footnote{https://github.com/BuzzFeedNews/2016-10-facebook-fact-check}
    
    \cite{potthast2018stylometric,newsveracity_santia@icwsm2018} & News published on Facebook from 9 agencies over a week close to the 2016 U.S. election. & \red{mostly false} (104), \red{mixture of true and false} (245), \blue{mostly true} (1669), no factual (264)  & Journalist experts from BuzzFeed. & 2,282  \\ \hline

    \textbf{Fact-Checked-Stat} \cite{vlachos2014fact} & Statements fact-checked from popular fact-checking websites labeled by journalists. & \blue{true} (32), \blue{mostly true} (34) , \red{half true} (68), \red{mostly false} (37), \red{false} (49), fiction (1)  & Journalists from fact-checking websites.  & 221 \\ \hline
 
    \textbf{Fake-News-Net}\footnote{https://github.com/KaiDMML/FakeNewsNet} \cite{shu2017fake} & A repository for an ongoing data collection project for fake news research including news content and social context features with reliable group truth fake news labels. & \red{fake} (211), \blue{real} (211)  & Journalist experts from BuzzFeed and fact-checkers from PolitiFact.com.  & 422 \\ \hline
 
    \textbf{Fake-Real-News}\footnote{https://github.com/GeorgeMcIntire/fake\_real\_news\_dataset} & News articles published during 2015-2016 along with their titles. The entire corpus is built crawling real news with New York Times\footnote{https://developer.nytimes.com/} and NPR\footnote{https://www.npr.org/api} APIs and fake news from Kaggle dataset items to ensure an uniform distribution of the samples from both the classes. & \red{fake} (3,164), \blue{real} (3,171)  & Journalists for true news and human annotators from BS Detector for fake news. & 6,335  \\  \hline
 
    \textbf{Fake-Satire} \cite{golbeck2018fake} & Dataset of fake news and satire that are hand-coded, verified, and, in the case of fake news, include rebutting stories.  & \red{fake news} (283), satire (203) & Researchers based on an article from a fact-checking site or a piece of information that disproves a claim.  & 486  \\ \hline
    
    \textbf{FA-KES} \cite{salem2019fa} & A fake news dataset around the Syrian war (i.e. reports on war incidents that took place from 2011 to 2018.) & \red{fake} (378), \blue{true} (426) & Semi-supervised fact-checking labeling approach. & 804 \\ \hline
 
    \textbf{Fake-Twitter-Science} \cite{vosoughi2018spread} & All of the verified true and false news distributed on Twitter from 2006 to 2017. The data comprise $\sim$126,000 instances (rumors cascades) tweeted by $\sim$3 million people more than 4.5 million times.  & \blue{true} (24,409), \red{false} (82,605), \red{mixed} (19,287)  & Agreement between fact-checkers from six independent fact-checking organizations. & 126,301 \\ \hline

    \textbf{Kaggle}\footnote{https://github.com/JasonKessler/fakeout} & Text and metadata from fake and biased news sources around the web from BS Detector\footnote{\url{http://bsdetector.tech/}}. & bias (443), bs (11,492), conspiracy (430), \red{fake} (19), hate (246), junksci (102), satire (146), state (121)  & Human annotators from BS Detector. & 12,997 \\ \hline
 
    \textbf{LIAR} \cite{wang2017liar} & Short statements from PolitiFact.com manually labeled. & \red{half-true} (2638), \red{false} (2511), \blue{mostly-true} (2466), \red{barely-true} (2108), \blue{true} (2063), \red{pants-fire} (1050) & Fact-checkers from PolitiFact.com. &  12,836  \\
    \hline
    \end{tabular}%
    \caption{Labeled datasets for fake news detection task.}
  \label{tab:datasets}%
\end{table*}%

At a high-level, these datasets of fact-checked content were labeled according to different scales, as either fake or true by expert journalists, fact-checking websites, and industry detectors\footnote{BS Detector: \url{https://gitlab.com/bs-detector/bs-detector}.},

providing different pieces of information and contexts that allow us to extract distinct types of features~\cite{shu2017fake}. We also can note that, except for the dataset related to the Syrian war, most of the related fake news datasets focus on US political news, entertainment news, or satire articles with extra information from traditional media like Twitter and Facebook. 

The dataset shared in this paper is complementary to those described in Table \ref{tab:datasets}, being unique for three main reasons. 
Firstly, it covers  fake news shared during the same type of major event (i.e., elections) but in two different occasions and contexts (i.e.,  the 2018 Brazilian elections and  the 2019 Indian elections). Thus, the dataset is not restricted to the bias of a single event.
Secondly, most currently publicly available datasets are restricted to textual content. Misinformation is not necessarily limited to text. Our dataset adds, for the first time, image based misinformation to the mix.
Thirdly, our dataset contains not only the labeled fact-checked images but also provides how such images were shared on WhatsApp in the groups we monitored. 

WhatsApp emerged as a key tool for spreading misinformation during the two analyzed elections. However, as WhatsApp is an encrypted and very closed network, it is hard to extract  information on the misinformation dissemination occurred through this messaging app. To the best of our knowledge, our dataset is the first of its kind to provide such broad and diverse information about WhatsApp.

\section{Dataset Construction}

\begin{figure*}[ht!] %[ht!]
	\centering
	   \includegraphics[width=0.9\linewidth]{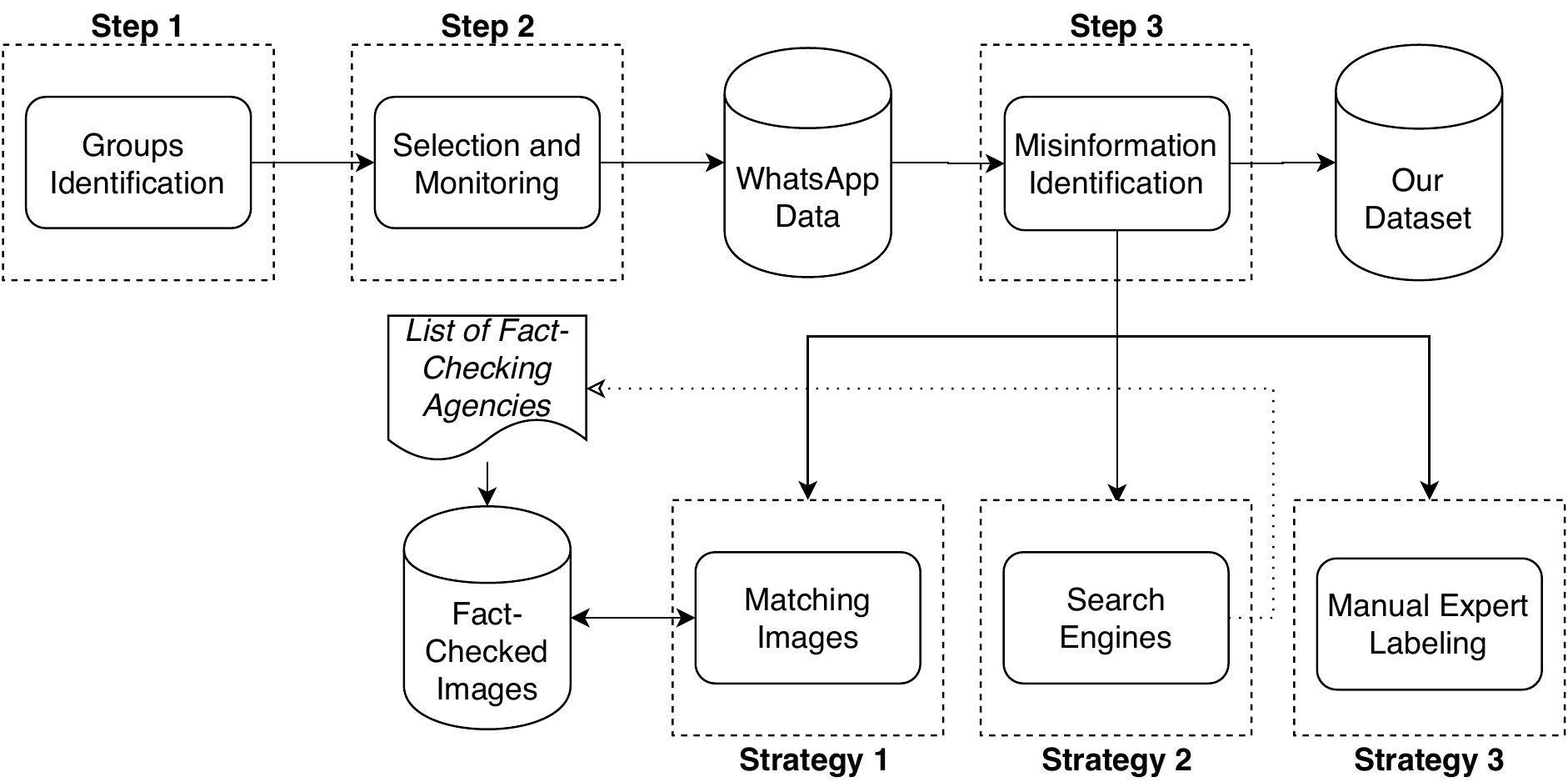}
	\caption{Overview of data collection.}
	\label{fig:data_collection}
\end{figure*}

To gather fact-checked images shared on WhatsApp during the Brazilian and Indian elections, the first step is collecting the data from WhatsApp. To do that, we followed the approach used by~\cite{garimella2018whatapp} and~\cite{resendeWWW19} to get access to messages posted on  WhatsApp groups. Their approach joins a set of publicly accessible groups  and collects all messages (including text, images, audios and videos) shared on the application.
Figure~\ref{fig:data_collection} presents an overview of our data collection process.

Given a set of invitation links to publicly accessible groups on WhatsApp (\textit{Step 1}), we automatically joined these groups and saved all data coming from them. We selected over 400 and 5,000 groups from Brazil and India, respectively, dedicated to political discussions which we monitored during the election period in these two countries, i.e. August -- November 2018 in Brazil, and October 2018 -- June 2019 in India.
We obtained invitation links to these groups by performing selected queries related to each election on Google search engine and Twitter.

In addition to exploring these groups which have been shown to be quite widely used in Brazil and India~\cite{lokniti2018,reuters2019report}, we choose to filter only messages containing images. 
Previous efforts showed that images are the most frequent type of media content, as well as an important source of misinformation~\cite{resendeWWW19}. 
Also, images are harder to manipulate and can be easily shared across groups and even platforms, unlike text, which is much easier to change during dissemination.
Thus, we monitored the selected groups during 2.5 month around each election (\textit{Step 2}) 
and, for each message collected, we filter out those containing images. This lead us to a set of over 34K and 810K images shared by more than 17k and 63K users on Brazil and India from which we extracted the following fields: 

\begin{itemize}
    \item \texttt{group\_id}: ID of the group the message was posted;
    \item \texttt{user\_id}: User ID\footnote{Note that we consider that these are, actually, phone numbers. Hence, different numbers by the same person are mapped to different identifiers.}; 
    \item \texttt{image\_id}: ID of the attached image file; 
    \item \texttt{cluster\_image\_name}: similarity identifier between images, i.e. all images with the same identifier -- cluster\_image\_name -- are the same. We will show more details of this process in the following sections. 
    \item \texttt{timestamp}: time record (i.e. date and hour) in which an image was sent (timestamp format).
\end{itemize}

We emphasize that all sensitive information (i.e., group names and phone numbers) were anonymized in order to ensure the privacy of users.
The dataset only consists of images that were publicly fact-checked and anonymized user/group information. Hence, to the best of our knowledge, it does not violate the WhatsApp terms of service.

In the next step, the images were clustered based on their similarity using a perceptual hashing technique from Facebook\footnote{\url{https://github.com/facebook/ThreatExchange/tree/master/hashing/pdq/}} to group together all images that are visually similar. 
This allows us to track the spread of an image (and minor variants of the image) through multiple groups on WhatsApp, including the time when an image was shared, the user who shared it and the group  where it was shared.
This is valuable information that allows us to assess not only the popularity of individual images but also how it was disseminated and its reach  within the WhatsApp groups.  It also aids us in the next task to create a set of unique images that need to be checked if they contain misinformation.
Since all the images in a cluster are visually similar, in our dataset, we only provide one image representative for each cluster, but we also include in the dataset additional information about all occurrences of an individual image sent through the groups.

\subsection{Misinformation Identification}

After tracking all images shared on WhatsApp and their dissemination, the next step is to identify those containing misinformation. This step (\textit{Step 3}) consists of identifying, among the images that circulated on WhatsApp during the monitored period, those that contained misinformation. We accomplish this task through three distinct approaches: (1) based on matching the images from WhatsApp to images that were already fact-checked by major online fact-checking agencies in Brazil and India
, (2) by making use of search engines, and (3) by manual expert labeling. 
These three complementary strategies help us to maximize the chances of finding misinformation images.

\subsubsection{Matching with Fact-Checked Data.}
First, we crawled all images which were fact-checked from popular fact-checking websites from Brazil\footnote{\url{aosfatos.org}, \url{www.e-farsas.com}, \url{veja.abril.com.br/blog/me-engana-que-eu-posto/},  \url{g1.globo.com/e-ou-nao-e/}, \url{piaui.folha.uol.com.br/lupa/}, and \url{www.boatos.org}} and India\footnote{\url{altnews.in},\url{boomlive.in},\url{smhoaxslayer.com},\url{factchecker.in},\url{factly.in},\url{fakenewscounter.com},\url{check4spam.com}}. 
For each of the fact-checking website, we developed a script  to parse and save all content and images that were fact-checked. For each post, when explicitly available, we also obtain the verdict of the fact check (fake or true).

In total, we collected over 100k fact-checked images from Brazil and about 20k images from  India.

Next, we matched this set of fact-checked images and those circulating on WhatsApp using a perceptual hashing approach that generates hashes to compare visually similar images.
We used the state-of-the-art technique developed and being used at Facebook: the PDQ hashing~\footnote{\url{https://github.com/facebook/ThreatExchange/blob/master/hashing/hashing.pdf}}.
This approach, an improvement over the commonly and broadly used pHash~\cite{zauner2010implementation}, can detect near visually similar images even if cropped differently or have small amounts of text overlaid on them. Facebook PDQ hash produces a 256 bit hash string using a discrete cosine transformation algorithm. 
We used the PDQ algorithm to compute the pairwise match between each fact-checked image collected and each image shared in our WhatsApp dataset. From all the images that matched with fact-checking websites, we only retained images that were labeled false by manually verifying the verdict given by the website.

\subsubsection{Search Engines.} 
In our second strategy, we automated the process of searching each image shared on the WhatsApp groups on the Web by using the Reverse Google Image search as proposed in~\cite{resendeWWW19}. Given the search results for an image, we checked whether any of the returned pages belong to one of the main fact-checking domains from Brazil and India, according to the previously defined set. If so, as performed at the first strategy, we also parsed the fact-checking page and automatically labeled the fact-checked image as fake or true depending on how the image was labeled on the fact-checking page. 

\subsubsection{Manual Expert Labeling.}
Finally, we also asked three journalists from India to label a sample of 2,500 images. 
The journalists are experts in fact-checking and regularly report on digital platforms.
The 2,500 images consist: (i) a random sample of 500 images taken from our 1.6 million images, and (ii) a set of 2,000 most shared images from our data.
We sampled these two sets differently in order to get a sense of the underlying distribution of misinformation in the rest of the dataset.
We built an annotation portal specifically for the journalists to view and annotate the images.
This gave us 605 unique misinformation images.
Note that this expert labeling was done only for India.

As shown in Table~\ref{tab:dataset} our final and shareable dataset contains 135 distinct fact-checked images from Brazil and 897 distinct fact-checked from India. Those images were shared by 919 different users in 184 different groups in Brazil and by over 12,000 users on 2,662 groups in India, being shared a total of 2,209 and 57,766 times in Brazil and India respectively.

\begin{table*}[h]
\centering
\caption{Overview of dataset on fact-checked images --labeled as fake-- shared in WhatsApp during the Brazilian and Indiam Elections.}
%\small
\label{tab:dataset}
\begin{tabular}{l|l|l|l|l|l|l|}
\cline{2-7}
                             & \multicolumn{1}{c|}{\#Users} & \multicolumn{1}{c|}{\#Groups} & \multicolumn{1}{c|}{\begin{tabular}[c]{@{}c@{}}Unique\\ Images\end{tabular}} & \multicolumn{1}{c|}{\begin{tabular}[c]{@{}c@{}}Total\\ Images\end{tabular}} & \multicolumn{1}{c|}{\#Shares} &  
                             \multicolumn{1}{c|}{Time Span} 
                             \\ \hline
\multicolumn{1}{|l|}{Brazil (MISINFO)} & 919                       & 184                           & 135                                                                       & 2,145                                                                       & 2,209  & 2018/08 - 2018/11     \\ \hline
\multicolumn{1}{|l|}{Brazil (NOTMISINFO)} &  439                      &   138                         &    15                                                                    &  829                                                                      & 850  & 2018/08 - 2018/11     \\ \hline
\multicolumn{1}{|l|}{Brazil (RANDOM)} & 1,104                        &     206                       & 304                                                                       & 2,148                                                                       & 2,205  & 2018/08 - 2018/11     \\ \hline \hline
\multicolumn{1}{|l|}{India (MISINFO)}  & 12,713                          & 2,662                            & 897                                                                         & 33,517                                                                 &  57,766  & 2018/07 - 2019/06     \\ \hline
\multicolumn{1}{|l|}{India (NOTMISINFO)}  & 12,558                          & 2,793                            & 765                                                                         &              27,201                                                    &  43,363  & 2018/07 - 2019/06     \\ \hline

\multicolumn{1}{|l|}{India (RANDOM)}  & 1,717                          & 1,452                            & 957                                                                         & 968                                                                 &  4,118  & 2018/07 - 2019/06     \\ \hline

\end{tabular}
\end{table*}

In addition to the dataset of misinformation images, we also provide two sets of roughly similar size.
(i) NOTMISINFO: A set of images manually checked by the three annotators to be true images. Note that this set of images mostly contains highly shared images which are not misinformation, and (ii) RANDOM: A set of randomly sampled images from all the images we collected.
Together with the images, we also provide details of users who shared them, groups in which they were shared and when they were shared.

We need to observe that there are some limitations regarding the data gathered in this work.
We used 3 different strategies in order to detect the most fact-checked images shared on WhatsApp 
but it is possible that some images containing misinformation and included in our dataset had not been checked by any of the fact-checking agencies used in this work, or were not properly matched using the hashing technique. 
Thus, we cannot comment on the recall  and the sampling bias present in the dataset.

Moreover, the groups monitored here are just a portion of the entire WhatsApp network. We cannot claim  statistical representativeness, as we do not have access to all groups on WhatsApp. 
Still, to date, this is the largest sample of WhatsApp  available for research.

We also believe that our dataset presents some strong advantages. First of all, it uses labels from fact-checking agencies, thus relying on specialist labeling. Also, an image-based dataset  on fake news  is not as common as a text-based dataset.
Furthermore, it covers an important context for studies on fake news around the world, namely elections, but it consider two distinct scenarios, and thus is not restricted to peculiarities of only one isolated event. 
Finally, the dataset explores the context of the closed network of WhatsApp and popular content circulated there. 
WhatsApp is becoming very important to studies on fake news, especially in Brazil and India, the countries which we gather data from. However, it is a challenging task to get any data from this encrypted messaging app due to the closed nature of its network. Thus, this dataset can provide a useful resource in understanding this phenomenon, especially in terms of the dissemination process.

\section{Conclusions}

In this work, we present a novel dataset of fact-checked images shared in WhatsApp during the Brazilian and India Elections can be useful for research in a variety of contexts. Here, we present a few possibilities for research directions using this dataset.

\subsubsection{Better Comprehension of the misinformation landscape.}
Countering misinformation is a typical adversarial fight. Every election, misinformation campaigns explore new ways to manipulate opinion and new defense mechanism are created aiming at least to mitigate the misinformation campaigns.
As an example, the efforts that attempt to understand the abuse of WhatsApp in the Brazilian elections have motivated countermeasures deployed in the Spanish elections\footnote{\url{https://elpais.com/elpais/2019/03/18/inenglish/1552900378_672737.html}, \url{https://www.independent.co.uk/news/world/europe/spain-elections-whatsapp-podemos-channel-close-left-ing-de-olmo-a8886481.html}} as well as worldwide changes in WhatsApp to slow down the dissemination of viral content\footnote{\url{https://www.theguardian.com/technology/2019/jan/21/whatsapp-limits-message-forwarding-fight-fake-news}}. By sharing our dataset, we hope other researchers can provide a better and faster comprehension of the phenomena, triggering new countermeasures in the upcoming elections.

\subsubsection{Identification of Invariants.} 
A key challenge to counter fake news is to identify its patterns which remain unchanged across contexts and dissemination environment. Our dataset offers data from two countries abused by misinformation campaigns inside WhatsApp, thus, offering novel contexts and an unexplored dissemination environment for comparisons with, for example, the 2016 USA elections. 

\subsubsection{Development of Automatic Detection Tools.} 
Our dataset can be used as ground truth to incorporate strategies based on machine learning algorithms to detect fake news automatically. Fact-checked images may contain unique signals that would allow one to automatically identify cases of manipulated images. We hope our dataset might allow the studies in this direction.  

\section{Acknowledgments}
\noindent This research was partially supported by Ministério Público de Minas Gerais (MPMG), project Analytical Capabilities, as well as grants from Fundação de Amparo à Pesquisa do Estado de Minas Gerais (FAPEMIG), Conselho Nacional de Desenvolvimento Científico e Tecnológico (CNPq), and Coordenação de Aperfeiçoamento de Pessoal de Nível Superior (CAPES).

\bibliographystyle{aaai}
\balance
\bibliography{references}

\begin{thebibliography}{}

\bibitem[\protect\citeauthoryear{Arun}{2019}]{arun2019whatsapp}
Arun, C.
\newblock 2019.
\newblock On whatsapp, rumours, and lynchings.
\newblock {\em Economic \& Political Weekly} 54(6):30--35.

\bibitem[\protect\citeauthoryear{Bessi and Ferrara}{2016}]{bessi2016social}
Bessi, A., and Ferrara, E.
\newblock 2016.
\newblock Social bots distort the 2016 us presidential election online
  discussion.
\newblock {\em First Monday} 21(11-7).

\bibitem[\protect\citeauthoryear{Garimella and
  Tyson}{2018}]{garimella2018whatapp}
Garimella, K., and Tyson, G.
\newblock 2018.
\newblock Whatapp doc? a first look at whatsapp public group data.
\newblock In {\em Proc. of the Int'l AAAI Conference on Weblogs and Social
  Media (ICWSM)}.

\bibitem[\protect\citeauthoryear{Golbeck \bgroup et al\mbox.\egroup
  }{2018}]{golbeck2018fake}
Golbeck, J.; Mauriello, M.; Auxier, B.; Bhanushali, K.~H.; Bonk, C.;
  Bouzaghrane, M.~A.; Buntain, C.; Chanduka, R.; Cheakalos, P.; Everett, J.~B.;
  et~al.
\newblock 2018.
\newblock Fake news vs satire: A dataset and analysis.
\newblock In {\em Proc. of the Int'l ACM Conference on Web Science
  (WebScience)}.

\bibitem[\protect\citeauthoryear{Hui \bgroup et al\mbox.\egroup
  }{2018}]{hui2018hoaxy}
Hui, P.-M.; Shao, C.; Flammini, A.; Menczer, F.; and Ciampaglia, G.~L.
\newblock 2018.
\newblock The hoaxy misinformation and fact-checking diffusion network.
\newblock In {\em Proc. of the Int'l AAAI Conference on Weblogs and Social
  Media (ICWSM)}.

\bibitem[\protect\citeauthoryear{Lokniti}{2018}]{lokniti2018}
Lokniti, C.
\newblock 2018.
\newblock How widespread is whatsapp's usage in india?
\newblock
  \url{https://www.livemint.com/Technology/O6DLmIibCCV5luEG9XuJWL/How-widespread-is-WhatsApps-usage-in-India.html}.

\bibitem[\protect\citeauthoryear{Melo \bgroup et al\mbox.\egroup
  }{2019}]{melo2019whatappMonitor}
Melo, P.; Messias, J.; Resende, G.; Garimella, K.; Almeida, J.; and Benevenuto,
  F.
\newblock 2019.
\newblock Whatsapp monitor: A fact-checking system for whatsapp.
\newblock In {\em Proc. of the Int'l AAAI Conference on Weblogs and Social
  Media (ICWSM)}.

\bibitem[\protect\citeauthoryear{Newman \bgroup et al\mbox.\egroup
  }{2019}]{reuters2019report}
Newman, N.; Fletcher, R.; Kalogeropoulos, A.; and Nielsen, R.~K.
\newblock 2019.
\newblock {Reuters Institute Digital News Report 2019 }.
\newblock Reuters Institute for the Study of Journalism.

\bibitem[\protect\citeauthoryear{N{\o}rregaard, Horne, and
  Adal{\i}}{2019}]{norregaard2019nela}
N{\o}rregaard, J.; Horne, B.~D.; and Adal{\i}, S.
\newblock 2019.
\newblock Nela-gt-2018: A large multi-labelled news dataset for the study of
  misinformation in news articles.
\newblock In {\em Proc. of the Int'l AAAI Conference on Weblogs and Social
  Media (ICWSM)}.

\bibitem[\protect\citeauthoryear{Potthast \bgroup et al\mbox.\egroup
  }{2018}]{potthast2018stylometric}
Potthast, M.; Kiesel, J.; Reinartz, K.; Bevendorff, J.; and Stein, B.
\newblock 2018.
\newblock A stylometric inquiry into hyperpartisan and fake news.
\newblock In {\em Proc. of the Annual Meeting of the Association for
  Computational Linguistics (ACL)}.

\bibitem[\protect\citeauthoryear{Resende \bgroup et al\mbox.\egroup
  }{2019}]{resendeWWW19}
Resende, G.; Melo, P.; Sousa, H.; Messias, J.; Vasconcelos, M.; Almeida, J.;
  and Benevenuto, F.
\newblock 2019.
\newblock (mis)information dissemination in whatsapp: Gathering, analyzing and
  countermeasures.
\newblock In {\em Proc. of the ACM Web Conference (WWW)}.

\bibitem[\protect\citeauthoryear{Ribeiro \bgroup et al\mbox.\egroup
  }{2019}]{ribeiro2019@fat}
Ribeiro, F.~N.; Saha, K.; Babaei, M.; Henrique, L.; Messias, J.; Benevenuto,
  F.; Goga, O.; Gummadi, K.~P.; and Redmiles, E.~M.
\newblock 2019.
\newblock On microtargeting socially divisive ads: A case study of
  russia-linked ad campaigns on facebook.
\newblock In {\em Proc. of the Conference on Fairness, Accountability, and
  Transparency (FAT)}.

\bibitem[\protect\citeauthoryear{Salem \bgroup et al\mbox.\egroup
  }{2019}]{salem2019fa}
Salem, F. K.~A.; Al~Feel, R.; Elbassuoni, S.; Jaber, M.; and Farah, M.
\newblock 2019.
\newblock Fa-kes: A fake news dataset around the syrian war.
\newblock In {\em Proc. of the Int'l AAAI Conference on Weblogs and Social
  Media (ICWSM)}.

\bibitem[\protect\citeauthoryear{Santia and
  Williams}{2018}]{newsveracity_santia@icwsm2018}
Santia, G., and Williams, J.
\newblock 2018.
\newblock Buzzface: A news veracity dataset with facebook user commentary and
  egos.
\newblock In {\em Proc. of the Int'l AAAI Conference on Weblogs and Social
  Media (ICWSM)}.

\bibitem[\protect\citeauthoryear{Shu \bgroup et al\mbox.\egroup
  }{2017}]{shu2017fake}
Shu, K.; Sliva, A.; Wang, S.; Tang, J.; and Liu, H.
\newblock 2017.
\newblock Fake news detection on social media: A data mining perspective.
\newblock {\em ACM SIGKDD Explorations Newsletter} 19(1):22--36.

\bibitem[\protect\citeauthoryear{Tardaguila, Benevenuto, and
  Ortellado}{2018}]{nyt2018benevenuto}
Tardaguila, C.; Benevenuto, F.; and Ortellado, P.
\newblock 2018.
\newblock Fake news is poisoning brazilian politics. whatsapp can stop it.
\newblock
  https://www.nytimes.com/2018/10/17/opinion/brazil-election-fake-news-whatsapp.html.

\bibitem[\protect\citeauthoryear{Vlachos and Riedel}{2014}]{vlachos2014fact}
Vlachos, A., and Riedel, S.
\newblock 2014.
\newblock Fact checking: Task definition and dataset construction.
\newblock In {\em Proc. of the ACL Workshop on Language Technologies and
  Computational Social Science}.

\bibitem[\protect\citeauthoryear{Vosoughi, Roy, and
  Aral}{2018}]{vosoughi2018spread}
Vosoughi, S.; Roy, D.; and Aral, S.
\newblock 2018.
\newblock The spread of true and false news online.
\newblock {\em Science} 359(6380):1146--1151.

\bibitem[\protect\citeauthoryear{Wang}{2017}]{wang2017liar}
Wang, W.~Y.
\newblock 2017.
\newblock "liar, liar pants on fire": A new benchmark dataset for fake news
  detection.
\newblock In {\em Proc. of the Annual Meeting of the Association for
  Computational Linguistics (ACL)}.

\bibitem[\protect\citeauthoryear{Wardle \bgroup et al\mbox.\egroup
  }{2019}]{firstdraf2019report}
Wardle, C.; Pimenta, A.; Conter, G.; and andPedro Burgos, N.~D.
\newblock 2019.
\newblock {Comprova: An Evaluation of the Impact of a Collaborative Journalism
  Project on Brazilian Journalists and Audiences}.
\newblock First Draft.

\bibitem[\protect\citeauthoryear{Zauner}{2010}]{zauner2010implementation}
Zauner, C.
\newblock 2010.
\newblock Implementation and benchmarking of perceptual image hash functions.

\end{thebibliography}
\balance
\end{document}